\documentstyle[12pt]{article}
\begin{document}
%

\topskip 2cm

\begin{center}
{\large\bf Order $\alpha_s^2$ Contributions to the Fragmentation Functions
in $e^+e^-$-Annihilation} \\
\vspace{2.5cm}
{\large W.L. van Neerven}
\footnote{talk presented at the Rencontres de 
Physique de La Valee d'Aoste, La Thuile, Italy, March 2-8, 1997}\\
\vspace{.5cm}
\centerline{\sl Instituut-Lorentz,}
\centerline{\sl University of Leiden,}
\centerline{\sl PO Box 9506, 2300 RA Leiden,}
\centerline{\sl The Netherlands.}
\vspace{1.5cm}
\vfil
\begin{abstract}
We discuss the most important results which are obtained from a recent
calculation of the order $\alpha_s^2$ corrections to the coefficient
functions contributing to the fragmentation functions $F_k(x,Q^2)$ 
($k=T,L,A$) in $e^+e^-$-annihilation. From these fragmentation functions
one can derive the corresponding integrated transverse ($\sigma_T$),
 longitudinal
($\sigma_L$) and asymmetric ($\sigma_A$) cross sections. The sum $\sigma_{tot} 
= \sigma_T+ \sigma_L$ corrected up to order $\alpha_s^2$ agrees with the well 
known result in the literature providing us with an independent check on our 
calculation. It turns out that the order $\alpha_s^2$ corrections to the
transverse and asymmetric quantities are small. However these corrections
to $F_L(x,Q^2)$ and $\sigma_L$ are large so that one gets a better agreement
between the theoretical predictions and the data obtained from the LEP 
experiments.
\end{abstract}
\end{center}
\newpage
\noindent
The measurement of the fragmentation functions in the process
\begin{equation}
  \label{eq1}
  e^+ + e^- \rightarrow \gamma,Z \rightarrow H + ``X",
\end{equation}
provides us, in addition to other experiments like deep inelastic lepton-hadron
scattering, with a new test of scaling violations as predicted by perturbative
quantum chromodynamics (QCD). 
Here $``X"$ denotes any inclusive final hadronic state and $H$ represents 
either a specific charged outgoing hadron or a sum over all charged hadron 
species. This process has been studied over a wide range of energies of many
different $e^+e^-$-colliders. The most recent data are coming from LEP (see
\cite{abreu},\cite{busk},\cite{akers}).
The unpolarized
differential cross section of the above process is given by 
\begin{equation}
  \label{eq2}
  \frac{d^2\sigma^H}{dx\,d\cos\theta} = \frac{3}{8}
(1+\cos^2\theta)\frac{d\sigma_T^H}{dx}
  + \frac{3}{4}\sin^2\theta\frac{d\sigma_L^H}{dx} + \frac{3}{4}\cos\theta
  \frac{d\sigma_A^H}{dx}.
\end{equation}
The Bj{\o}rken scaling variable $x$ is defined by
\begin{equation}
  \label{eq3}
  x = \frac{2pq}{Q^2},\hspace{8mm} q^2 = Q^2 > 0,\hspace{8mm} 0 < x \leq 1,
\end{equation}
where $p$ and $q$ are the four-momenta of the produced particle $H$ and the 
virtual vector
boson ($\gamma$, $Z$) respectively. In the centre of mass (CM) frame of the 
electron-positron
pair the variable $x$ can be interpreted as a fraction of the beam energy 
carried away
by the hadron $H$. The variable $\theta$ denotes the angle of emission of 
particle $H$
with respect to the electron beam direction in the CM frame. 
The transverse, longitudinal
and asymmetric cross sections in (\ref{eq2}) are defined by $\sigma_T^H$, 
$\sigma_L^H$, and
$\sigma_A^H$ respectively. The latter only shows up if the intermediate vector
boson is given
by the $Z$-boson and is absent in purely electromagnetic annihilation.\\
Before the advent of LEP1 the CM energies were so low ($\sqrt s<M_Z$) that
$\sigma_A$ could not be measured and no effort was made to separate $\sigma_T$
{}from $\sigma_L$ so that only data for the sum of the above cross sections in
 eq. (\ref{eq2}) were available.
Recently after LEP1 came into operation ALEPH and OPAL obtained data for
$\sigma_T$ and $\sigma_L$ separately and the latter collaboration even made a 
measurement of $\sigma_A$ for the first time. 
The separation of $\sigma_T$ and $\sigma_L$
is important because the latter cross section enables us to extract the strong
coupling constant $\alpha_s$ and allows us to determine the gluon fragmentation
density $D_g^H(x,\mu^2))$ with a much higher accuracy as could be done before.
Furthermore the measurement of $\sigma_A$ provides us with information
on hadronization effects since the QCD corrections are very small.
In the QCD improved parton model, which describes the production of the parton 
denoted by $l$ ($l=q,\bar q,g$)
and its subsequent fragmentation into hadron $H$, the cross sections 
$\sigma_k^H$ ($k=T,L,A$)
can be expressed as follows
\begin{eqnarray}
  \label{eq4}
  \lefteqn{\frac{d\sigma_k^H}{dx} = \int_x^1\,\frac{dz}{z}\Biggl[
    \sigma_{\rm tot}^{(0)}(Q^2)\,\Biggl \{
    D_S^H\left(\frac{x}{z},\mu^2\right)
    {\cal C}_{k,q}^S(z,Q^2/\mu^2)
    + D_g^H\left(\frac{x}{z},\mu^2\right)\cdot}
\nonumber\\[2ex]
  && {}\cdot{\cal C}_{k,g}^S (z,Q^2/\mu^2)\Biggr \}
  +\sum_{f=1}^{n_f}\,\sigma_f^{(0)}(Q^2)\,
  D_{NS,f}^H\left(\frac{x}{z},\mu^2\right)
  {\cal C}_{k,q}^{NS}(z,Q^2/\mu^2)\Biggr],
\end{eqnarray}
for $k=T,L$. In the case of the asymmetric cross section we have
\begin{equation}
  \label{eq5}
  \frac{d\sigma_A^H}{dx} = \int_x^1\,\frac{dz}{z}\Biggl[\sum_{f=1}^{n_f}\,
  A_f^{(0)}(Q^2)
  D_{A,f}^H\left(\frac{x}{z},\mu^2\right)
  {\cal C}_{A,q}^{NS}(z,Q^2/\mu^2)\Biggr].
\end{equation}
In the formulae (\ref{eq4}) and (\ref{eq5}) we have introduced the following 
notations.
The function $D_l^H(z,\mu^2)$ denotes the parton fragmentation density 
corresponding to the hadron of species $H$.
Further we have defined the singlet (S) and non-singlet (NS, A) combinations of 
the quark fragmentation densities. They are given by
\begin{eqnarray}
D_S^H(z,\mu^2) & = & \frac{1}{n_f}\,\sum_{q=1}^{n_f}\left(D_q^H(z,\mu^2) +
    D_{\bar{q}}^H(z,\mu^2)\right) ,\label{eq6}\\
D_{NS,q}^H(z,\mu^2) &=& D_q^H(z,\mu^2) + D_{\bar{q}}^H(z,\mu^2) 
- D_S^H(z,\mu^2),\label{eq7}\\
D_{A,q}^H(z,\mu^2)& =& D_q^H(z,\mu^2) - D_{\bar{q}}^H(z,\mu^2).\label{eq8}
\end{eqnarray}
The index $q$ stands for the quark species and $n_f$ denotes the number of 
light flavours.
Further the variable $\mu$
appearing in $D_l^H(z,\mu^2)$ and the coefficient functions 
${\cal C}_{k,l}(z,Q^2/\mu^2)$ ($k=T,L,A$; $l=q,g$)
stands for the mass factorization scale which for
convenience has been put equal to the renormalization scale. 
The pointlike cross
section $\sigma_q^{(0)}$ and the asymmetry factor $A_q^{(0)}$ of the process 
$e^+ + e^- \rightarrow q + \bar{q}$ can be found in eq. (10) and eq. (15) of
\cite{rn1} respectively. The total cross section, summed over all flavours is
given by $\sigma_{\rm tot}^{(0)}(Q^2) = \sum_{q=1}^{n_f}\sigma_q^{(0)}(Q^2)$.
The QCD corrections in eqs. (\ref{eq4}), (\ref{eq5}) are described by the 
coefficient functions ${\cal C}_{k,l}^r$
($k=T,L,A$; $l=q,g$) which can be distinguished with respect to the flavour 
group
$SU(n_f)$ into a singlet ($r=S$) and a non-singlet part ($r=NS$).
The gluonic coefficient function only receives contributions from flavour 
singlet channel partonic
subprocesses so that we can drop the superscript $S$ on ${\cal C}_{k,g}$. 
However the quark
coefficient functions get flavour singlet as well as flavour non-singlet 
contributions.
Up to first order in the strong coupling constant $\alpha_s$ it turns out that
${\cal C}_{k,q}^{NS}={\cal C}_{k,q}^S$. However in higher 
order both quantities
start to deviate from each other. Hence we define the purely singlet 
coefficient function
${\cal C}_{k,q}^{PS}$ via
\begin{equation}
  \label{eq9}
  {\cal C}_{k,q}^S 
= {\cal C}_{k,q}^{NS} + {\cal C}_{k,q}^{PS}.
\end{equation}
Like ${\cal C}_{k,g}$ the purely singlet coefficient function only receives 
contributions
{}from the flavour singlet channel partonic subprocesses which in the former
case show up for the first time in order $\alpha_s^2$.\\
Using charge conjugation invariance of the strong interactions
one can show that ${\cal C}_{A,q}^{NS} = -{\cal C}_{A,\bar{q}}^{NS}$
and ${\cal C}_{A,q}^{PS} = {\cal C}_{A,g} = 0$. This implies
that to $\sigma_A^H$ in eq. (\ref{eq5}) only non-singlet channel partonic 
subprocesses can contribute. Another important property of the coefficient 
function is that they do not depend on the probe $\gamma$ or $Z$
so that one can extract the overall pointlike cross section 
$\sigma_q^{(0)}$
or the asymmetry factor $A_q^{(0)}$ . 
However this is only
true if all quark masses are equal to zero and if one sums over all 
quark members in one
family provided the latter appear in the inclusive state of the partonic 
subprocess (see \cite{rn1})
From (\ref{eq2}) we can derive the total hadronic cross section
\begin{equation}
\label{eq10}
\sigma_{\rm tot}(Q^2) = \frac{1}{2}\sum_H\,\int_0^1dx\,\int_{-1}^1 d\cos\theta\,
\left(x\frac{d^2\sigma^H}{dx\,d\cos\theta}\right) = \sigma_T(Q^2) + 
\sigma_L(Q^2),
\end{equation}
with
\begin{equation}
\label{eq11}
\sigma_k(Q^2) = \frac{1}{2}\sum_H\,\int_0^1dx\,x\frac{d\sigma_k^H}{dx},
\hspace{4mm} (k=T,L,A),
\end{equation}
where one has summed over all types of outgoing hadrons $H$.
From the momentum conservation sum rule given by
\begin{equation}
\label{eq12}
\sum_H\,\int_0^1 dx x D_l^H(x,\mu^2) = 1 \hspace*{1cm} \mbox{$l=q,\bar q,g$}
\end{equation}
and eqs. (\ref{eq4}), (\ref{eq11}) one can derive
\begin{equation} 
\label{eq13}
\sigma_k(Q^2) = \sigma_{\rm tot}^{(0)}(Q^2)\int_0^1dx\,x\Biggl[
{\cal C}_{k,q}(x,Q^2/\mu^2)^S+ \frac{1}{2}{\cal C}_{k,g}(x,Q^2/\mu^2)\Biggr],
\end{equation}
Finally we also define the transverse, longitudinal and asymmetric 
fragmentation functions
$F_k^H(x,Q^2)$\footnote{Notice that we make a distinction in nomenclature 
between
the fragmentation densities $D_q^H$,$D_g^H$ and the fragmentation functions 
$F_k^H$.}
\begin{equation}
\label{eq14}
F_k^H(x,Q^2) = \frac{1}{\sigma_{\rm tot}^{(0)}(Q^2)}\,\frac{d\sigma_k^H}{dx},
\hspace{4mm} k=(T,L,A).
\end{equation}
One observes that
the above fragmentation functions
are just the timelike analogues of the structure functions measured in for deep 
inelastic electron-proton scattering.
The calculation of the coefficient functions corrected up to order $\alpha_s^2$
proceeds in the same way as done for the Drell-Yan process in \cite{hnm} and
for deep inelastic lepton-hadron scattering in \cite{zn}. Denoting the
intermediate vector bosons $\gamma$ and $Z$ by the symbol $V$ we have the
following parton subprocesses. In zeroth order we have the Born reaction
\begin{equation}
\label{eq15}
V \rightarrow "q" + \bar q
\end{equation} 
where $"l"$ ($l=q,\bar q,g$) denotes the detected parton which subsequently
fragments into the hadron of species $H$.
In next-to-leading order (NLO) one has to include the one-loop virtual  
corrections to reaction (\ref{eq15}) and the parton subprocesses
\begin{eqnarray}
V &\rightarrow & "q" + \bar q + g\label{eq16}\\
V &\rightarrow & "g" + q + \bar q\label{eq17} .
\end{eqnarray}
After mass factorization of the collinear divergences which arise in the above
processes one obtains the coefficient functions which are presented in
\cite{aemp},\cite{bf},\cite{nw}.
The determination of the order $\alpha_s^2$ contributions involves the 
computation of the two-loop corrections to (\ref{eq15}) and the one-loop  
corrections to eqs. (\ref{eq16}),(\ref{eq17}). Furthermore one has to calculate
 the following subprocesses
\begin{eqnarray}
V &\rightarrow & "q" + \bar q + g + g\label{eq18}\\
V &\rightarrow & "g" + q + \bar q + g\label{eq19}\\
V &\rightarrow & "q" + \bar q + q + \bar q\label{eq20}. 
\end{eqnarray}
In reaction (\ref{eq20})) the two anti-quarks, which are inclusive, 
can be identical as
well as non identical. Notice that in the above reactions the detected quark 
can be replaced by the detected anti-quark so that in reaction (\ref{eq20}) one 
can also distinguish between the final states containing identical quarks and
non identical quarks. After mass factorization and renormalization for which 
we have chosen the ${\overline {\rm MS}}$-scheme one obtains the order
$\alpha_s^2$ contributions to the coefficient functions which are presented in
\cite{rn1},\cite{rn2},\cite{rn3}.
The most important results of our calculations can be summarized as follows.
From eq. (\ref{eq13}) and the coefficient functions originating from the 
processes above
we can obtain $\sigma_L$ and $\sigma_T$ corrected up to order $\alpha_s^2$
\begin{eqnarray}
\label{eq21}
\sigma_L(Q^2)& = & \sigma^{(0)}_{\rm tot}(Q^2)\Biggl[\frac{\alpha_s(\mu^2)}{
4\pi}
C_F\Biggl\{3\Biggr\} + \left(\frac{\alpha_s(\mu^2)}{4\pi}\right)^2\Biggl[
    C_F^2\Biggl\{-\frac{15}{2}\Biggr\} + C_AC_F\Biggl\{ \nonumber\\[2ex]
  && {} -11 \ln\frac{Q^2}{\mu^2}
  -\frac{24}{5}\zeta(3) + \frac{2023}{30}\Biggr\} + n_fC_FT_f\Biggl\{
  4\ln\frac{Q^2}{\mu^2} - \frac{74}{3}\Biggr\}\Biggr]\Biggr],
\nonumber\\
\end{eqnarray}
\begin{eqnarray}
\label{eq22}
\sigma_T(Q^2)& = & \sigma^{(0)}_{\rm tot}(Q^2)\Biggl[ 1 + \left(
\frac{\alpha_s(\mu^2)}{4\pi}\right)^2\Biggl[ C_F^2\Biggl\{6\Biggr\} 
+ C_AC_F\Biggl\{
    -\frac{196}{5}\zeta(3) - \frac{178}{30}\Biggr\} \nonumber\\[2ex]
  && {}+ n_fC_FT_f\Biggl\{ 16\zeta(3) + \frac{8}{3}\Biggr\}\Biggr]\Biggr].
\end{eqnarray}
Addition of $\sigma_L$ and $\sigma_T$ yields the well known answer for
$\sigma_{\rm tot}$ (\ref{eq10}) which is in agreement with the literature 
\cite{ckt}.
Hence this quantity provides us with a check on our calculation of the
longitudinal and transverse coefficient functions. Notice that in lowest order
$\sigma_{\rm tot}$ only receives a contribution from the transverse cross
section whereas the order $\alpha_s$ contribution can be only attributed
to the longitudinal part. In order $\alpha_s^2$ both $\sigma_L$
and $\sigma_T$ contribute to $\sigma_{\rm tot}$.\\
Because of the high sensitivity of expression (\ref{eq21}) to the value 
of $\alpha_s$,
the longitudinal cross section provides us with an excellent tool to measure the
running coupling constant.\\
To illustrate the sensitivity of $\sigma_L$ on $\alpha_s$ 
we have plotted in Fig. 1 in LO and NLO the ratio
\begin{equation}
\label{eq26}
R_L(Q^2)=\frac{\sigma_L(Q^2)}{\sigma_{tot}(Q^2)}
\end{equation}
as a function of $Q$ (CM-energy of the $e^+e^-$ system). 
Our input parameters for the running coupling constant are 
$\Lambda_{LO}^{(5)} = 108~MeV$ ($\alpha_s^{LO}(M_Z) = 0.122$) and
$\Lambda_{\overline {\rm MS}}^{(5)} = 227~MeV$ ($\alpha_s^{NLO}(M_Z) = 0.118$). 
Fig. 1 reveals that the order $\alpha_s$ corrections are
appreciable and they vary from 48\% ($Q=10~GeV$) down to 28\% ($Q=200~GeV$)
with respect to the LO result. Furthermore one observes an improvement of the
renormalization scale dependence while going from $R_L^{LO}$ to $R_L^{NLO}$.
Comparing with the experimental value $R_L = 0.057\pm 0.005$, measured by 
OPAL \cite{akers},
one observes a considerable improvement when the order
$\alpha_s^2$ contributions are included. However there is still a little 
discrepancy between $R_L^{NLO}$ at $\mu=Q=M_Z$, and the
data. This can either be removed by choosing a larger 
$\Lambda_{\overline {\rm MS}}$ or by
including the masses of the heavy quarks $c$ and $b$ in the calculation of the
coefficient functions. Also a contribution of higher twist effects can maybe not
neglected (see \cite{nw}).

We now want to investigate the effect of the order $\alpha_s^2$ contributions
to the longitudinal and transverse fragmentation functions 
$F_L(x,Q^2)$ and $F_T(x,Q^2)$ eq. (\ref{eq14}). Here we have summed over the 
following hadron species i.e. $H=\pi^{\pm},K^{\pm},p,\bar p$.
Choosing the parametrization of the fragmentation densities in \cite{bkk}, with
$\Lambda_{LO}^{(5)} = \Lambda_{\overline {\rm MS}}^{(5)} = 190~MeV$, we have
plotted $F_L^{LO}$ and $F_L^{NLO}$ in Fig. 1. We observe
that $F_L^{LO}$ is below the data in particular in the small $x$-region. The
agreement with the data becomes better when the order $\alpha_s$ 
corrections are
included although at very small $x$ $F_L^{NLO}$ is still smaller than the values
given by experiment. This figure reveals the importance of the higher order
corrections to $F_L(x,Q^2)$ for
the determination of the fragmentation densities. We have also shown results
for $F_T^{NLO}$ and $F_T^{NNLO}$ in Fig. 3 using the same fragmentation density
set. Both fragmentation functions agree with the data except that $F_T^{NNLO}$
gets a little bit worse at very small $x$. 
Furthermore $F_T^{NLO}$ and $F_T^{NNLO}$ hardly
differ from each other which means that the order $\alpha_s^2$ 
corrections are small.
One also notices that $F_L$ constitutes the smallest part of the
total fragmentation function $F=F_T+F_L$ which can be inferred from Figs. 2 and
3. This in particular holds at large $x$ where $F_T >> F_L$. Hence a
fit of the fragmentation densities to the data of $F_T$ is not
sufficient to give a precise prediction for $F_L$ and one
has to include the data of the longitudinal part to provide us with a better set
of fragmentation densities. This in particular holds for $D_g^H(z,\mu^2)$ 
in the small $z$-region for which the order $\alpha_s^2$ contribution to $F_L$
will be needed. The latter will certainly change the parametrization of the
gluon fragmentation density given by ALEPH in \cite{busk} and OPAL in 
\cite{akers}.\\
%

{\large \bf Figure Captions}\\

\begin{description}

\item[Fig. 1] 
The ratio $R_L=\sigma_T/\sigma_{\rm tot}$. Dotted lines:
$R_L^{LO}$; solid lines: $R_L^{NLO}$. Lower curve: $\mu=2Q$;
middle curve: $\mu=Q$; upper curve: $\mu=Q/2$. The data point at 
$Q=M_Z$ is from OPAL \cite{akers}.
\item[Fig. 2]
The longitudinal fragmentation function $F_L(x,Q^2)$ at $\mu=Q=M_Z$.
Dotted line: $F_L^{LO}$; solid line: $F_L^{NLO}$. The data are from
ALEPH \cite{busk} and OPAL \cite{akers}. The fragmentation
density set is from \cite{bkk}.
\item[Fig. 3]
The transverse fragmentation function $F_T(x,Q^2)$ at $\mu=Q=M_Z$.
Dotted line: $F_T^{NLO}$; solid line: $F_T^{NNLO}$. The data are from
ALEPH \cite{busk} and OPAL \cite{akers}. The fragmentation
density set is from \cite{bkk}.
\end{description}

%
\renewcommand{\thesection}{}
\setlength{\unitlength}{1cm}
\begin{figure}[p]
  \begin{center}
    \input figure1
  \end{center}
  \caption{\label{fig:f1}}
\end{figure}
\begin{figure}[p]
  \begin{center}
    \input figure2
  \end{center}
  \caption{\label{fig:f2}}
\end{figure}
\begin{figure}[p]
  \begin{center}
    \input figure3
  \end{center}
  \caption{\label{fig:f3}}
\end{figure}

\begin{thebibliography}{99}
\bibitem{abreu}
P. Abreu et al. (DELPHI), Phys. Lett. 311B (1993) 408.
%
\bibitem{busk}
D. Buskulic et al. (ALEPH), Phys. Lett. 357B (1995) 487.
\bibitem{akers}
R. Akers et al. (OPAL), Z. Phys. C68 (1995) 203.
%
\bibitem{rn1}
P.J. Rijken and W.L. van Neerven, Nucl. Phys. 487B (1997) 233.
%
\bibitem{hnm}
R. Hamberg, W.L. van Neerven, T. Matsuura, Nucl. Phys. B359 (1991) 343.
%
\bibitem{zn}
E.B. Zijlstra and  W.L. van Neerven, Nucl. Phys. B383 (1992) 525.
%
\bibitem{aemp}
P. G. Altarrelli, R.K. Ellis, G. Martinelli, S.-Y. Pi, Nucl. Phys. B160 (1979)
301.
%
\bibitem{bf}
R. Baier and K. Fey, Z. Phys. C2 (1979) 339.
%
\bibitem{nw}
P. Nason and B.R. Webber, Nucl. Phys. B421 (1993) 473.
%
\bibitem{rn2}
P.J. Rijken and W.L. van Neerven, Phys. Lett. B386 (1996) 422.
%
\bibitem{rn3}
P.J. Rijken and W.L. van Neerven, Phys. Lett. 392B (1997) 207.
%
\bibitem{ckt}
K.G. Chetyrkin, A.L. Kataev, F.V. Tkachov, Phys. Lett. 85B (1979) 277;\\
M. Dine and J. Sapirstein, Phys. Rev. Lett. 43 (1979) 668;\\
W. Celmaster and R.J. Gonsalves, Phys. Rev. Lett. 44 (1980) 560.
%
\bibitem{bkk}
J. Binnewies, B.A. Kniehl, G. Kramer, Z. Phys. C65 (1995) 471.
%
\vspace*{1cm}
\end{thebibliography}
\end{document}